\documentclass[aps,twocolumn,showpacs,amsmath,amssymb,superscriptaddress]{revtex4}
\usepackage{graphicx}
\usepackage{amssymb,amsmath}
\usepackage{dcolumn}
\usepackage{bm}
\begin{document}

\title{Ultrafast pump-probe dynamics in ZnSe-based semiconductor quantum-wells}

\author{Henni Ouerdane}
\author{George Papageorgiou}
\author{Ian Galbraith}
\author{Ajoy K. Kar}
\author{Brian S. Wherrett}
\affiliation{Physics Department, Heriot-Watt University, Edinburgh EH14 4AS, UK}
\email{I.Galbraith@hw.ac.uk}

\begin{abstract}
Pump-probe experiments are used as a controllable way to investigate the properties of photoexcited semiconductors, in particular, the absorption saturation. We present an experiment-theory comparison for ZnSe quantum wells, investigating the energy renormalization and bleaching of the excitonic resonances. Experiments were performed with spin-selective excitation and above-bandgap pumping. The model, based on the semiconductor Bloch equations in the screened Hartree-Fock approximation, takes various scattering processes into account phenomenologically. Comparing numerical results with available experimental data, we explain the experimental results and find that the electron spin-flip occurs on a time scale of 30 ps.
\end{abstract}

\pacs{190.0190, 190.7110, 190.5970, 320.0320, 320.7130.}

\maketitle

\section {Introduction}

From an experimental point of view, one can investigate the optical properties of semiconductors by exciting carriers (by means of optical pumping or carrier injection) and measuring the absorption of a subsequent probe pulse. By comparison of this spectrum with the linear absorption spectrum, one obtains information on the influence of the excitation on the absorption phenomenon and insight into the electronic and optical properties of the electron-hole plasma. Interpretation of experimental results is, however, nontrivial, given the substantial influence of Coulomb and many-body effects, which give rise to a rich variety of broadening and energy renormalizations. Moreover, the time evolution of the initial electron–hole plasma makes the whole problem challenging, both theoretically and numerically. In this paper we present a model that can describe the time evolution of the nonequilibrium electron-hole system but that is also simple enough to account for many dynamical processes that occur, using different polarizations of the pump and probe beams.

Much previous theoretical work has focused on the study the absorption phenomenon in semiconductor quantum-wells in a quasi-equilibrium situation \cite{SCH86, HAU93, CHO99, BIN91, SNE94}. Here, based on the available experimental data, we move beyond such a quasi-equilibrium situation. We include six dynamical processes that lead eventually to a thermal quasi-equilibrium in the electron-hole plasma: relaxation of the hot carriers' distributions toward Fermi-Dirac distributions, thermalization among the carrier gases, plasma cooling, carrier spin-flip, scattering between the light- and heavy-hole bands, and recombination (both radiative and nonradiative). A true microscopic treatment accounting for all these many-body effects would be computationally prohibitive. Instead, we use a phenomenological approach to describe the time evolution of the hot electron-hole plasma. The absorption spectra are evaluated from the time-dependent semiconductor Bloch equations (SBE) in the screened Hartree-Fock approximation. In this work, as we focus on time scales over many picoseconds, we do not need to consider the full coherent dynamics involving the nonlinear scattering of pump light into the probe direction as a result of the nonlinear polarization interaction. The focus we choose allows and justifies the phenomenological treatment of the scattering processes for our qualitative analysis. Varying the delay between the pump and probe beams will allow us to obtain the time evolution of both the bleaching of the exciton peaks and the energy renormalizations, which, compared to experimental data will give an estimation of the time scale of the scattering processes mentioned above. Simultaneously, as we show below, the dynamics of the electron-hole plasma (density, temperature, plasma screening and distribution of each type of carrier in each spin state) can be monitored.

The aim of this paper is to present our model for the time evolution of the electron-hole plasma created by spin-selective excitation in the absorption continuum and study its influence on absorption spectra. The experimental setup and results are described in Section 2. In Section 3, we present our theoretical model for the time evolution of the electron-hole plasma, including the semiconductor Bloch equations that have to be solved numerically together with the rate equations used. We discuss our numerical results, comparing them with experimental data, in Section 4.

\section{Ultrafast pump and probe experiments}

A femtosecond laser system consisting of a Beamlok argon-ion ($\mbox{Ar}^+$) laser, a Tsunami mode-locked Ti:Sapphire laser, a Merlin Q-switched Nd:YLF laser, a Spitfire pulsed Ti:Sapphire regenerative amplifier and an ultrafast kilohertz optical parametric amplifier (OPA), was used for the generation of the ultrafast pump pulses. The $\mbox{Ar}^+$ laser and the Merlin laser were the excitation sources for the Tsunami laser and the Spitfire amplifier respectively. The Tsunami output was fed to the Spitfire where it was temporally stretched, amplified, and finally temporally compressed. The Spitfire output provided the pump beam for the frequency conversion processes in the OPA. The overall system was capable of delivering a 1kHz train of $\sim$ 150 fs pulses and the wavelength was tuned at 459 nm ($\sim$ 2.69 eV). For the generation of the white light continuum, $\sim$ 5\% of the Spitfire output ($\lambda$ = 800 nm) was focused on a 10-mm-thick quartz cuvette containing deionized water. The pump pulse was used to excite the semiconductor sample, which was mounted on the cryostat and cooled at 4K. The pump power, and therefore the carrier density, could be controlled by the use of a neutral density filter. The pump pulse power incident in the cryostat was 0.06 mW. The changes induced in the transmitted probe pulse energy were measured by an optical spectrum multichannel
analyzer as a function of the time delay between the pump and the probe pulses. Using a glass microscope slide to monitor its stability we selected a small portion of the white-light continuum before it fell onto the sample. The spot size radius of the probe beam was 190 $\mu$m, and it was considerably smaller than that of the pump in order to probe a region of uniform photoexcited density. Both pump and probe beams were circularly polarized and independently controllable by $\lambda$/4 plates. Opposite circular (OCP) and same circular polarization (SCP) configurations were employed. We provided the time resolution by delaying the white light continuum pulses relative to the pump pulses. The experimental work was performed on a ZnSe/ZnCdSe multiple quantum well structure of twenty 4nm-wide wells grown by molecular beam epitaxy on GaAs substrate. The 20\% Cd content in the wells produces light-hole--heavy-hole exciton splitting of more than 30 meV.

The absorption spectra, Fig.~1, show that at early times, both the heavy-hole and light-hole exciton peaks are bleached but not shifted much. The broadening that is due to the interaction-induced dephasing plays an important role in both SCP and OCP, but for SCP the heavy-hole exciton peak is more bleached because of the Pauli blocking effect that reduces the oscillator strength. In contrast, the light hole exciton peak is more bleached in OCP. A detailed description and an interpretation of these experimental results are given below with our numerical analysis.

\begin{figure}[!rh]
\centering
\scalebox{0.33}{\includegraphics*{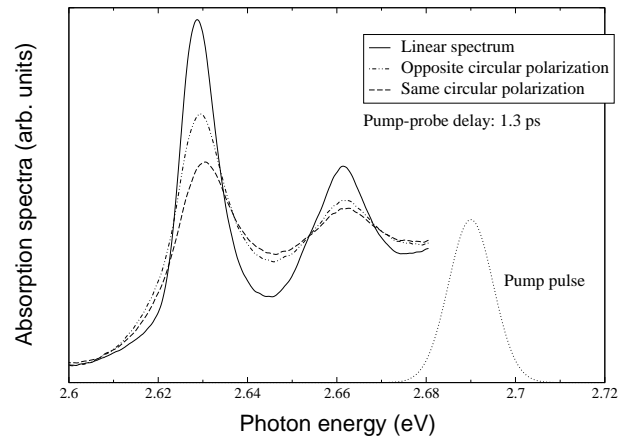}}
\caption{Measured absorption spectra. Comparison of OCP SCP absorption.}
\end{figure}

Increasing the delay between the pump and the probe beams suggests that many dynamical processes occur in the electron-hole plasma and change the shape of the absorption spectra. The dynamics of the absorption spectra is shown in Fig. 2, where both the bleaching ($\Delta\alpha/\alpha$) and the energy shift of the heavy-hole exciton peak are given as functions of the delay between the pump and the probe. Experimental data show an overall decay of the exciton peak bleaching as well as a convergence of the OCP and the SCP curves. They also show an initial blueshift at early times and an energy shift that brings the resonances back to the linear spectrum exciton resonance. The energy shift exhibits the same type of behavior as the exciton bleaching: The OCP and SCP curves converge on a 30-ps time scale.

\begin{figure}[!rh]
\centering
\scalebox{0.33}{\includegraphics*{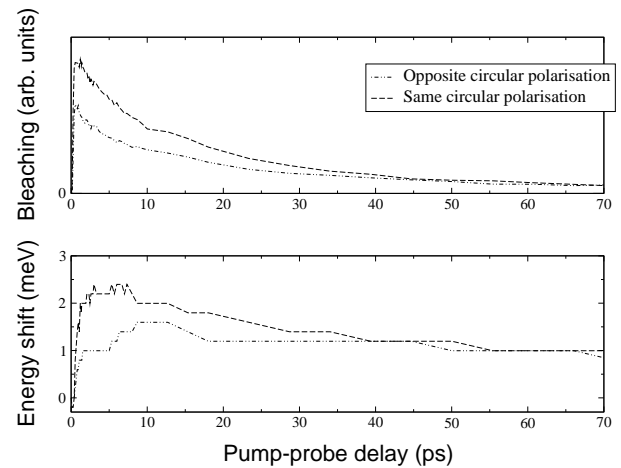}}
\caption{Measured absorption spectra dynamics. Comparison of OCP and SCP heavy-hole exciton peaks' bleaching and shift.}
\end{figure}

\section{Theoretical model}

To describe and explain what we observe, we constructed a theoretical model that describes the time evolution of the electron-hole plasma and its influence on the absorption spectra.

\subsection{Polarization dynamics}

Inasmuch as the heavy-hole--light-hole band splitting is $\Delta E_{\rm cs}$ = 30 meV, we neglect the heavy-hole (hh) and light-hole (lh) coupling. The interband polarization equations \cite{LIN88} are:

\begin {eqnarray}\label{eq1}
\hbar\frac{\displaystyle \partial}{\displaystyle \partial t}~ p_{\bm k}^{\lambda} (t) & = & - i (e_{{\rm e},k} + e_{\lambda,k})~ p_{\bm k}^{\lambda} (t)
\nonumber\\ 
&&{}- i (n_{{\rm e},{\bm k}}^{\sigma} (t) + n_{\lambda,{\bm k}}^{\sigma'} (t) -1)~ \hbar\omega_{{\rm R},{\bm k}}^{\lambda} (t)~,
\end {eqnarray}

\noindent where the Rabi energies, $\hbar\omega_{{\rm R},{\bm k}}^{\lambda}$, are given by:

\begin {equation} \label{eq2}
       \hbar\omega_{{\rm R},{\bm k}}^{\lambda} (t) = d_{\rm cv}^{\lambda}~ {\mathcal E} (t) + \sum_{{\bm q}\ne{\bm k}} V_{|{\bm k}-{\bm q}|}^s~p_{{\bm q}}^{\lambda} (t)~,
\end {equation}

\noindent for $\lambda =$ hh, lh. The energies $e_{i,k}$ are the renormalized energies evaluated in the static plasmon-pole approximation including the contribution of the pair continuum \cite{HAU93}:

\begin {equation} \label{eq3}
  e_{i,k} = \epsilon_{i,k} + \Sigma_{{\rm exc},i}(k) + \frac{\displaystyle 1}{\displaystyle 2}~\Delta E_{\rm CH}~, ~~~~ i = \mbox{e}, \mbox{hh}~ \mbox{and}~ \mbox{lh},
\end {equation}

\noindent with $\Sigma_{{\rm exc},i}$ the screened exchange self-energy, $\Delta E_{\rm CH}$ the Coulomb hole energy \cite{ELL89}, and $n_{c,{\bm k}}^{\sigma} (t)$ are the occupancy of the carrier of type $c$ with spin $\sigma$ at the time $t$.

The temporal envelope of the probe field, ${\mathcal E} (t)$, is assumed to be Gaussian and the optical suceptibility $\chi (\omega)$ is defined as:

\begin {equation} \label{eq4}
  \chi (\omega) = \frac {\displaystyle {\hat P}(\omega)}{\displaystyle \epsilon{\hat {\mathcal E}}(\omega)},
\end {equation}

\noindent where $\epsilon$ is the dielectric constant and ${\hat P}(\omega)$ and ${\hat {\mathcal E}}(\omega)$ are the Fourier transform of the polarization function $P(t) = d_{\rm cv}\sum_{\bm k} p_k (t)$ and of the electric field ${\mathcal E}(t)$. It is a complex function whose imaginary part is proportional to the absorption: $\alpha (\omega) \propto \mbox{Im} \chi (\omega)$ \cite{HAU93}.

To solve Eq.~({\ref{eq1}) one needs knowledge of the carrier distributions to evaluate the phase space filling factor and the plasma temperature and density to calculate the screened Coulomb potential energy $V_{\bm q}^s$ entering the definition of the Rabi frequency, Eq.~(\ref{eq2}). We neglect coherent polarization nonlinearities, as we are considering above bandgap pumping and time scales longer than the dephasing time.

\subsection{Evolution of the carriers distributions}

In our model the time evolution of the distribution function, $n_{c,{\bm k}}^{\sigma}(t)$, of a carrier $c$ with a spin $\sigma$ takes several dynamical processes into account. Relaxation of the carrier distributions, carrier spin-flip, recombination and light hole scattering yield the following system of six coupled differential equations ($\sigma =~\uparrow$ or $\sigma =~\downarrow$, indicating the two spin-states):

\begin{widetext}
$$
    \frac {\displaystyle {\rm d}n_{{\rm e},{\bm k}}^{\sigma}} {\displaystyle {\rm d}t} = \frac {\displaystyle n_{{\rm e},{\bm k}}^{{\rm eq},\sigma} - n_{{\rm e},{\bm k}}^{\sigma}} {\displaystyle \tau_{\rm eq}^{\rm e}} + \frac {\displaystyle n_{{\rm e},{\bm k}}^{{\rm therm},\sigma} - n_{{\rm e},{\bm k}}^{\sigma}} {\displaystyle \tau_{\rm therm}^{\rm e}} + \frac {\displaystyle n_{{\rm e},{\bm k}}^{\sigma'} - n_{{\rm e},{\bm k}}^{\sigma}} {\displaystyle \tau_{\rm sf}^{\rm e}}~ + ~\frac {\displaystyle n_{{\rm e},{\bm k}}^{\sigma}~n_{{\rm hh},{\bm k}}^{\sigma'}}{\displaystyle \tau_{\rm rad}^{\rm e,hh}}~ - ~\frac {\displaystyle n_{{\rm e},{\bm k}}^{\sigma}~n_{{\rm lh},{\bm k}}^{\sigma}} {\displaystyle \tau_{\rm rad}^{\rm e,lh}} - ~\frac {\displaystyle n_{{\rm e},{\bm k}}^{\sigma}}{\displaystyle \tau_{\rm nr}^{\rm e}},
$$

$$
    \frac {\displaystyle {\rm d}n_{{\rm hh},{\bm k}}^{\sigma}} {\displaystyle {\rm d}t} = \frac {\displaystyle n_{{\rm hh},{\bm k}}^{{\rm eq},\sigma} - n_{{\rm hh},{\bm k}}^{\sigma}} {\displaystyle \tau_{\rm eq}^{\rm hh}} + \frac {\displaystyle n_{{\rm hh},{\bm k}}^{{\rm therm},\sigma} - n_{{\rm hh},{\bm k}}^{\sigma}} {\displaystyle \tau_{\rm therm}^{\rm hh}} + \frac {\displaystyle n_{{\rm hh},{\bm k}}^{\sigma'} - n_{{\rm hh},{\bm k}}^{\sigma}} {\displaystyle \tau_{\rm sf}^{\rm hh}}~ - ~\frac {\displaystyle n_{{\rm hh},{\bm k}}^{\sigma}~n_{{\rm e},{\bm k}}^{\sigma'}}{\displaystyle \tau_{{\rm rad}}^{\rm e,hh}} - ~\frac {\displaystyle n_{{\rm hh},{\bm k}}^{\sigma}}{\displaystyle \tau_{\rm nr}^{\rm hh}} + \frac {\displaystyle n_{{\rm lh},{\bm k}}^{\sigma}}{\displaystyle 2\tau_{\rm lh}} + \frac {\displaystyle n_{{\rm lh},{\bm k}}^{\sigma'}}{\displaystyle 2\tau_{\rm lh}},
$$

\begin {equation} \label{eq5}
    \frac{\displaystyle {\rm d}n_{{\rm lh},{\bm k}}^{\sigma}}{\displaystyle {\rm d}t} = \frac{\displaystyle n_{{\rm lh},{\bm k}}^{{\rm eq},\sigma} - n_{{\rm lh},{\bm k}}^{\sigma}} {\displaystyle \tau_{\rm eq}^{\rm lh}} + \frac {\displaystyle n_{{\rm lh},{\bm k}}^{{\rm therm},\sigma} - n_{{\rm lh},{\bm k}}^{\sigma}} {\displaystyle \tau_{\rm therm}^{\rm lh}} + \frac {\displaystyle n_{{\rm lh},{\bm k}}^{\sigma'} - n_{{\rm lh},{\bm k}}^{\sigma}} {\displaystyle \tau_{\rm sf}^{\rm lh}}~ - ~\frac {\displaystyle n_{{\rm lh},{\bm k}}^{\sigma}~n_{{\rm e},{\bm k}}^{\sigma}}{\displaystyle \tau_{\rm rad}^{\rm e,lh}} - ~\frac {\displaystyle n_{{\rm lh},{\bm k}}^{\sigma}}{\displaystyle \tau_{\rm nr}^{\rm lh}} - ~\frac {\displaystyle n_{{\rm lh},{\bm k}}^{\sigma}}{\displaystyle \tau_{\rm lh}}.
\end {equation}
\end{widetext}

These equations are numerically solved to yield the time dependence of the carrier distributions $n_{c,{\bm k}}^{\sigma}(t)$ whose values are 0--1. The various terms that enter into the above system are now described.\\

{ \bf Intraband scattering ($\tau_{\rm eq}$ terms)}: The optical pumping creates a population of hot carriers. One of the fastest processes (i.e. subpicosecond time scale \cite{SHA92}) that occurs in each band is the rapid equilibration of these carriers: due to carrier-carrier scattering the initial hot carrier distributions evolve towards Fermi--Dirac quasi-equilibrium distributions. Extensive work on this specific topic involving the quantum Boltzmann equation can be found in the literature \cite{BIN92, SCO92, JAH95_1, JAH95_2}. Here we use a phenomenological approach to describe the time evolution of the hot carrier distributions which is characterized by a relaxation time $\tau_{\rm eq}$ associated with intraband scattering. Thus, the quantities $n_{c,{\bm k}}^{{\rm eq},\sigma}$ in Eq.~(\ref{eq5}) are Fermi--Dirac distributions describing the quasi-equilibrium for each spin-polarized subsystem, which has the same carrier density and energy as the nonequilibrium distribution. Note that the intraband scattering does not change the carrier densities nor the total kinetic energy.\\

{\bf  Carrier thermalization ($\tau_{\rm therm}$ terms)}: A process that also influences the time evolution of the carrier distributions is the thermalization process among carriers of different types. The scattering between electrons and heavy- and light-holes is a process that drives the initial carrier temperatures, $T_c^{\sigma}$, to a common quasi-equilibrium temperature, $T_{\rm eq}$, which can be different from the lattice temperature, $T_{\rm lat}$. To evaluate $T_{\rm eq}$, one needs to calculate the total plasma energy $E_{\rm tot}$ and then compute the corresponding temperature $T_{\rm eq}$ assuming a Fermi--Dirac distribution. To account for the thermalization process, we suppose that the time evolution of the distribution functions is characterized by a phenomenological time $\tau_{\rm therm}$. Thus, the quantities $n_{c,{\bm k}}^{{\rm therm},\sigma}$ in Eq.~(\ref{eq5}) are Fermi--Dirac distributions that describe the quasi-equilibrium for each spin-polarized subsystem; they have the same carrier density but an energy corresponding to the common quasi-equilibrium temperature $T_{\rm eq}$.\\

{\bf Carrier spin-flip ($\tau_{\rm sf}$ terms)}: The spin-flip is a process by which the spin orientation of a carrier is reversed. Models that describe such a phenonemon have been proposed, e.g. Elliot-Yaffet \cite{ELL54, YAF63} and D'Yakonov-Perel \cite{DYA71}, but the detailed mechanisms responsible for the spin-flip process are not yet well understood despite extensive studies \cite{DAM91, FER91, POT99}. In this work, we only consider the spin-flip phenomenologically with the associated characteristic time $\tau_{\rm sf}$ that is of interest to us. The spin-flip introduces a coupling between the spin-states $\sigma$ and $\sigma'$ for a given type of carrier $c$, which considerably complicates the solution of Eq.~(\ref{eq5}). The value of $\tau_{\rm sf}$ is poorly known and one of our aims using this model will be to extract this value from the experimental data.\\

{\bf Recombination ($\tau_{\rm rad}$ and $\tau_{\rm nr}$ terms)}: We distinguish here between radiative and nonradiative recombinations. Radiative recombination is a process that occurs on relatively long time scale compared to the others described above ($\tau_{\rm rad}$ = 1.6 ns in ZnSe). The total observed luminescence intensity $I_{\rm pl}(t)$ is directly linked to the distribution functions of these carriers:

\begin {equation} \label{eq6}
 I_{\rm pl}^{\lambda}(t) \propto \sum_{{\bm k}} ~|d_{\rm cv}^{\lambda}|^2~n_{{\rm e},{\bm k}}^{\sigma}(t)~n_{\lambda,{\bm k}}^{\sigma'}(t)~,
\end {equation}

\noindent for the heavy and light hole optical transitions, where $\lambda =$ lh or hh, and $d_{\rm cv}$ is the dipole matrix element.

For a detailed study of radiative recombination and spontaneous emssion rate, see Ref.~\cite{KIR99}. Detailed calculations with which to evaluate $\tau_{\rm rad}$ can be found in Ref.~\cite{ROS98}. The term $\tau_{\rm nr}$ in the right hand side of Eq.~(\ref{eq5}) accounts for the nonradiative recombination which is a faster process than the radiative recombination at high plasma density.\\

{\bf Heavy hole and light hole scattering ($\tau_{\rm lh}$ terms)}: Away from zone center, both heavy and light holes are mixtures of the bulk valence band states. This mixture enhances the intersubband scattering between the heavy- and light-hole bands. Many processes that involve other quasi-particles (heavy and light holes, phonons, excitons \ldots) can facilitate this type of scattering, and it is nontrivial to assess the relative importance of each of the processes. To avoid such complication we make a simple approximation, assuming that this scattering is spin independent, e.g., that $|3/2, 1/2\rangle_{\rm lh}$ scatters equally into $|3/2, 3/2\rangle_{\rm hh}$. When the quasi-equilibrium between the heavy-hole and the light-hole bands is reached, their chemical potentials have to satisfy the relation $\mu_{\rm lh} = \mu_{\rm hh} - \Delta E_{\rm cs}$ because of the band splitting that is due to confinement and strain. Solving this equation at low temperature and for typical electron–hole plasma densities, i.e. of the order of $10^{11} \mbox{cm}^{-2}$ or more, yields a negligible light-hole density: $N_{\rm lh}/N \approx 0$. Hence we include a simple decay characterized by the time $\tau_{\rm lh}$ in Eqs. (5) to model the light-hole scattering.

\subsection{Plasma cooling}

Intraband scattering dominates the fast carrier distribution relaxation but is not a process that dissipates energy. Thus the initial electron–hole plasma temperature is determined by the kinetic energy of the nonequilibrium distribution created by the femtosecond pump pulse. Depending on the energy of the excitation pulse, the effective plasma temperature can be well above the lattice temperature. The most important source of energy dissipation is the coupling of the electronic system with the lattice. The plasma cooling can be treated by solution of the quantum Boltzmann equation that describes the carrier-phonon scattering \cite{JAH95_1, JAH95_2}, but we restrict ourselves to a phenomenological approach. Hence, the loss of carriers' kinetic energy obeys a rate equation:

\begin {equation} \label{eq9}
  \frac{\displaystyle {\rm d}E_{{\rm tot},c}^{\sigma}}{\displaystyle {\rm d}t} = \frac{\displaystyle E_{{\rm tot},c}^{{\rm eq},\sigma} - E_{{\rm tot},c}^{\sigma}}{\displaystyle \tau_{\rm cool}}~,
\end {equation}

\noindent where $E_{{\rm tot},c}^{{\rm eq},\sigma}$ is the total energy per unit area calculated from the quasi-equilibrium distribution $n_{c,E}^{{\rm eq},\sigma}$ at the lattice temperature. Eq.~(\ref{eq9}) has to be solved numerically.

To evaluate the effective carrier temperatures at each point in time, we use the time evolution of the nonequilibrium energies $E_{{\rm tot},c}^{\sigma}(t)$. We calculate first the quasi-equilibrium energies as an explicit function of temperature $E_c^{{\rm eq},\sigma}(T_c^{\sigma})$; then, equating $E_c^{{\rm eq},\sigma}(T_c^{\sigma})$ and the nonequilibrium energies $E_{{\rm tot},c}^{\sigma}(t)$ gives the effective temperature $T_c^{\sigma}$ at the time $t$:

\begin {eqnarray}\label{eq10}
E_c^{{\rm eq},\sigma}(T_c^{\sigma}) & = & \frac{\displaystyle m_c}{\displaystyle 2\pi\hbar^2}~ \left(e^{2\pi\hbar^2\beta_c^{\sigma}N_c^{\sigma}/m_c} - 1\right)
\nonumber\\
&&{}\times \int_0^{\infty} \frac{\displaystyle E ~dE}{\displaystyle e^{\beta_c^{\sigma}E} + e^{2\pi\hbar^2\beta_c^{\sigma}N_c^{\sigma}/m_c} - 1}~,
\end {eqnarray}

\noindent where $\beta_c^{\sigma} = 1/k_BT_c^{\sigma}$.

Knowledge of the distributions $n_{c,{\bm k}}^{\sigma}(t)$ and temperatures $T_c^{\sigma}$ allows us to evaluate the time evolution of the Pauli blocking factor that enters into the equations of motion for the interband polarizations, Eq.(\ref{eq1}).

\subsection{Time evolution of the plasma screening}

As the plasma temperature and the carrier densities evolve, the plasma screening changes. In this paper we are concerned with the time evolution of the electron-hole plasma and we need to find a simple way to evaluate it in a nonequilibrium situation. This Lindhard formula, which is valid for both equilibrium and nonequilibrium situations \cite{HAU93}, is our starting point:

\begin {equation} \label{eq12}
       \epsilon_q(\omega) = 1 - V_q~ \sum_{\sigma,{\bm k}}\sum_c~ \frac{\displaystyle n_{c,{\bm k}-{\bm q}}^{\sigma} - n_{c,{\bm k}}^{\sigma}}{\displaystyle \hbar(\omega - i\delta) + e_{c,{k-q}}^{\sigma} - e_{c,k}^{\sigma}}~.
\end {equation}

\noindent For ease is notation we do not explicitly denote the time dependence of the various physical quantities defined above. We have to simplify Eq.~(\ref{eq12}) as it is not practical to use that equation for numerical purposes because of its continuum of poles. We choose to work in the long wavelength limit $q \rightarrow 0$ \cite{HAU93}. Using a nonequilibrium distribution function we find:

\begin {equation} \label{eq13}
   \epsilon(q \rightarrow 0, \omega) = 1- V_q~ \frac{\displaystyle q^2}{\displaystyle \omega^2}\sum_{c,\sigma}~ \frac{\displaystyle  N_c^{\sigma}}{\displaystyle m_c}~,
\end {equation}

\noindent which shows that the time dependence of the distribution functions $n_{c,{\bm k}}^{\sigma}$ due to the relaxation process does not affect the expression of $\epsilon(q \rightarrow 0, \omega)$: it has the same form as in quasi-equilibrium calculations. So, to treat the nonequilibrium problem for the screening, we can calculate the screening from the quasi-equilibrium formulas that one can find in Refs.~ \cite{HAU93, ZIM88}. The time dependence of $\epsilon(q \rightarrow 0, \omega)$ is contained in $N_c^{\sigma}$. To obtain analytical results for the plasma screening we make use of the static plasmon-pole approximation \cite{HAU93}, and the screened Coulomb potential given by:

\begin {equation} \label{eq14}
  V_q^s = \frac{\displaystyle V_q}{\displaystyle \epsilon_q}~.
\end {equation}

\noindent  Note that our definition of the bare 2D Coulomb potential in a quantum well includes the form factor $f_q$ to account for the finite well width $w$ \cite{CHO99}.

\subsection{Initial conditions for the electron/hole plasma}

We now turn to the three pump-probe polarization configurations and the initial carrier distributions they produce as the starting point for the solution of Eqs.~(\ref{eq5}): 

\begin{enumerate}
  \item OCPs, for which the pump and the probe beams are both circularly polarized but in opposite senses,
  \item SCPs, for which the pump and the probe beams are cicularly polarized in the same sens,
  \item SLPs, for which the pump and the probe beams are both linearly senses.
\end{enumerate}

\begin{figure}[!rh]
\centering
\scalebox{0.33}{\includegraphics*{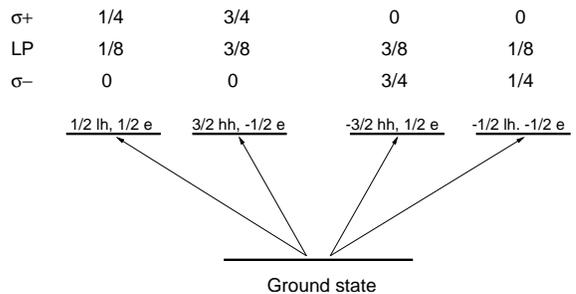}}
\caption{Selection rules for optical transitions achieved with polarized pump light}
\end{figure}

Depending on the pump-probe polarization configuration, we generate different couplings between the ground state and the various spin-states $|J,m_J\rangle$. In our analysis we include the two-fold degenerate conduction band $|1/2, 1/2\rangle_e$ and $|1/2, -1/2\rangle_e$, the two-fold degenerate heavy hole band $|3/2, 3/2\rangle_{hh}$ and $|3/2, -3/2\rangle_{hh}$ and the two-fold degenerate light hole band $|3/2, 1/2\rangle_{lh}$ and $|3/2, -1/2\rangle_{lh}$. The selection rules for zinc blende semiconductors are used \cite{BAS88} and the relative populations generated by optical pumping in the continuum are as depicted in Fig.~3. The ratio of 3 between the populations created from the heavy hole and light hole transitions comes from the ratio of the dipole matrix elements which describe the relative strengths of these optical transitions \cite{BAS88}.

In principle, the carrier dynamics during the pump process could be included in our model. However, since $\tau_{\rm eq} \sim$ 100 fs and we are not interested in the early coherent regime, we simply assume an initial carrier density with nonthermal Gaussian distributions reflecting the spectral width and the location of the pump pulse. This assumption has no significant impact on our numerical results as $\tau_{\rm eq}$ is of the order of 100 fs and saves considerable computational effort. In the SLP situation, electrons with a given spin $\sigma$ are created from both the light- and heavy-hole transitions. So, the initial distribution for the spin-polarized electron gases in the SLP is the sum of two Gaussian distributions, each centered appropriately.

\section{Comparison between numerical and experimental results}

To present and discuss the dynamics of the absorption spectra, it is useful first to describe the time evolution of the electron-hole plasma, which is influenced by the dynamical processes mentioned above. We are concerned with the interplay among the various dynamical processes included in our model and their influence on the time evolution of the carrier density, plasma energy and temperature. situations, namely, that when excitations have been performed by
circularly polarized light and that when they have been performed by linearly polarized light. The case of circular
polarization (CP) can describe both OCP and SCP situations: The dynamics of the carrier gas is the same; only
the interaction with the probe field is different.

The lattice temperature is taken to be $T_{\rm lat}$ = 77 K (we find in materials with strong Coulomb effects such as ZnSe, that strongly degenerate cases at 4 K are numerically prohibitive owing to the large number of $k$-states required near the Fermi energy) and the initial plasma density $N = 3 \times 10^{11} \mbox{cm}^{-2}$. The effective masses we use are $m_{\rm e} = m_{\rm lh} = 0.15 m_0$ and $m_{\rm hh} = 0.6 m_0$, where $m_0$ is the mass of the free electron. The dielectric constant is $\epsilon = 8.8$ and the bandgap $E_{\rm g} = 2.66$ eV. The phenomenological parameters entering Eqs.~(\ref{eq5}) are: $\tau_{\rm eq}$ = 0.1 ps, $\tau_{\rm cool}$ = 1 ps, $\tau_{\rm therm}$ = 1 ps. These characteristic times only give an order of magnitude and are taken from Ref.~\cite{SHA92}. The radiative recombination time $\tau_{\rm rad}$ =  1.6 ns is calculated for ZnSe parameters. We assume the nonradiative recombination to be density independent with $\tau_{\rm nr}$ = 30 ps taken from experiments, by fitting numerical results to experimental results (see Figs.~2 and 8). Inasmuch as we are pumping in the continuum, the excited valence band states are of a mixed heavy-hole--light-hole character and hence scatter efficiently. We chose $\tau_{\rm lh}$ = 0.5 ps for the light hole density decay. The spin-flip times, $\tau_{\rm sf}$ = 30 ps, have been chosen also comparing numerical results with experimental data. Various values for $\tau_{\rm nr}$ and $\tau_{\rm sf}$ have been tried numerically and we estimate the margin of error for the parameters to be $\pm$5 ps.

\subsection{Time evolution of the carrier densities}

First we study the time evolution of the carrier densities obtained by solution of Eqs.~(\ref{eq5}) and shown in Fig.~4.\\

\begin{figure}[!rh]
\centering
\scalebox{0.33}{\includegraphics*{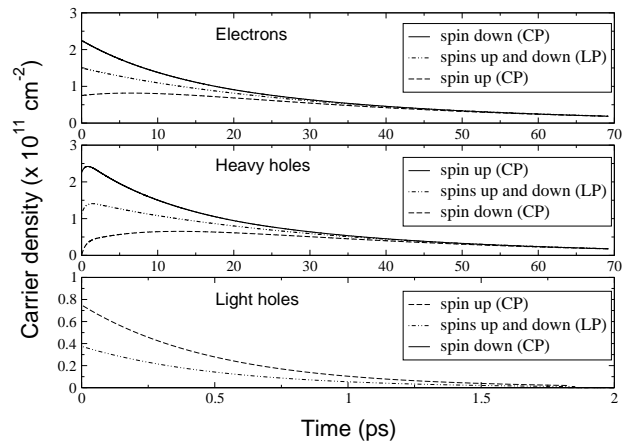}}
\caption{Time evolution of the spin-polarized gas populations ($10^{11} \mbox{cm}^{-2}$) created by spin-selective excitation. Note the short time scale for the light hole density decay.}
\end{figure}

{\bf Electrons}: The processes responsible for the change in population are the electron spin-flip and the recombination. The population of spin down electrons decreases while the population of electrons with opposite spin increases. The carrier spin-flip is a process that tends to create equal spin populations over a time scale given by $\tau_{\rm sf}^{\rm e} = 30$ ps. Then, the recombination which occurs on the same time scale given by $\tau_{\rm nr} = 30$ ps becomes dominant and drives both populations towards zero. That is the reason why the initially less populated electron gas exhibits a maximum at early times (10 ps). Note that for SLP, the spin-flip process has no effect on the dynamics of the electron densities. Hence, the only contributor to the density decay in the SLP situation is recombination. Also note that the radiative recombination is too slow (1.6 ns) to have much influence on the fast population dynamics.\\

{\bf Heavy holes}: Similar comments apply for the heavy hole population dynamics as above with $\tau_{\rm sf}^{\rm hh} = 30$ ps. However, the fast intersubband scattering that drives the light holes into the heavy hole band on a time scale given by $\tau_{\rm lh} = 0.5$ ps has to be considered here. That explains the fast initial rise of both heavy hole densities, before they start decaying. Note that the initial spin down heavy hole density is zero, unlike the electrons whose populations in both spin states is non-zero from the beginning because of optical transitions from both the heavy- and light-hole bands.\\

{\bf Light holes}: The time evolution of the light hole populations can also be described similarly as the electron populations with $\tau_{\rm sf}^{\rm lh} = 30$ ps. However, as in the case of the heavy holes, one has to consider the intersubband scattering. As it is a very fast process, the population of light holes decreases on a very short time scale given by $\tau_{lh}$ = 0.5 ps. After one picosecond the light hole density is negligible (but non-zero as heavy holes still scatter into the light hole band). Note that in the case of the spin down light hole gas, the population always remains very low: the spin-flip time is comparatively too long to make any significant change.

\subsection{Time evolution of the plasma energy}

The time evolution of the plasma energy is displayed on Fig.~5.\\

\begin{figure}[!rh]
\centering
\scalebox{0.33}{\includegraphics*{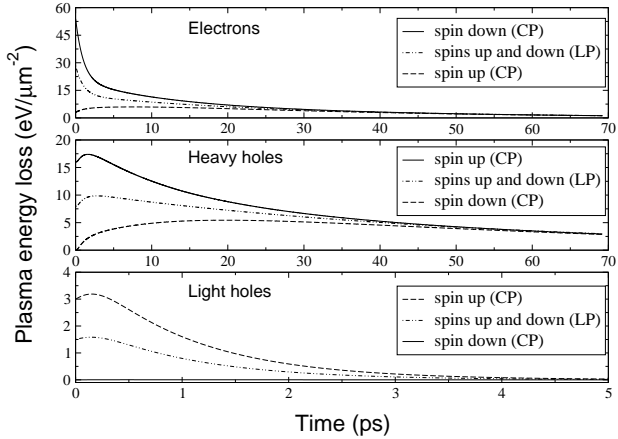}}
\caption{Time evolution of the spin-polarized gases energy loss ($\mbox{eV}/\mu\mbox{m}^2$). Note the very short time scale for the light hole gases' energy loss.}
\end{figure}

{\bf Electrons}: The initial total energy of the spin down electron gas in the CP case corresponds to a temperature above the lattice temperature and hence is a decreasing function of time because of the cooling process that occurs on a time scale given by $\tau_{\rm cool}^e = 1$ ps. One can observe that the energy of the electron gas with opposite spin is an increasing function of time before cooling down to the lattice energy. It is due to the thermalization process between carriers of different types and also to the fact that the density increases over the same amount of time because of the spin-flip , and hence increases the total energy. In the LP case, we only observe a cooling as the spin-flip has no influence on the time evolution of the densities.\\

{\bf Heavy holes}: The initial rise of the carrier densities for both heavy-hole gases as a result of light-hole scattering influences the behavior of the energy, which is an increasing function of time at early times. The thermalization process is also responsible for this rise, as the initial hole gases' energies are below the spin-down electron gas energy. In the case of the spin-down heavy-hole gas the increase occurs on a much longer time scale than for the opposite spin heavy-hole gas. This is the result of the spin–flip process that increases the population for the spin-up heavy-hole gas in CP.\\

{\bf Light holes}: The initial rise of the energy for the light hole gases is only due to the thermalization process. As the light hole density decay is fast, it only takes a few picoseconds for the light hole gas energy to become negligible. The population of the spin down light hole gas remains very low; so does the energy.

\subsection{Time evolution of the plasma temperature}

The effective temperatures of each carrier gas, shown in Fig.~6, are computed solving Eq.~(\ref{eq10}).\\

\begin{figure}[!rh]
\centering
\scalebox{0.33}{\includegraphics*{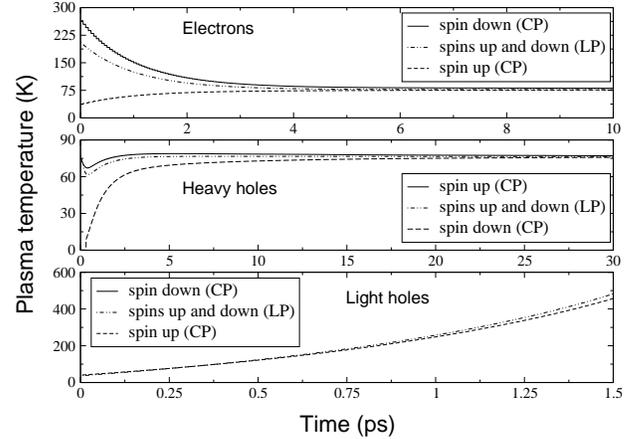}}
\caption{Time evolution of the spin-polarized gases' temperatures (K). Note the very short time scale for the light hole temperature evolution.}
\end{figure}

{\bf Electrons}: The time evolution of the electron temperatures follows exactly the time evolution of the electron gas energies, but it converges fast toward the lattice temperature, whereas the electron energies keep decreasing. This behavior is due to the fact that the electron densities also decrease by recombination, keeping the average electron kinetic energy constant.\\

{\bf Heavy holes}: The heavy-hole gas energies increase at early times; so do the densities, because of the light-hole scattering. In CP, for the spin-down heavy-hole gas, despite this increase in energy, the average heavy-hole kinetic energy decreases. Hence the effective temperature decreases. Then, because of the thermalization with the electrons and the lattice, the temperature starts increasing. For the spin-up heavy-hole gas we observe a monotonic increase toward the lattice temperature as it gains energy from the thermalization processes with the other carriers and the lattice. However, as the densities starts decreasing, the energy of the heavy holes also starts decreasing. But, as in the case of electrons, there is compensation between the energy loss and the density decay that makes the average heavy-hole kinetic energy constant when it reaches the lattice temperature.\\

{\bf Light holes}: The light-hole density decay is so fast compared to the plasma cooling that the average light-hole kinetic energy keeps increasing and the light-hole gases' temperatures go beyond the lattice temperature. In fact, the light-hole gases have no time to reach quasi-equilibrium with the lattice. We stop calculating the light-hole gases' effective temperatures when their population is small enough and their temperature high enough to have no influence on the absorption spectra (after 1.5 ps).

\subsection{Time evolution of the absorption spectra}

In this section we present numerical solution of Eqs.~(\ref{eq1}) and (\ref{eq5}) and discuss the behavior of the excitonic peaks' bleaching as well as their energy shift. First, we compare the calculated absorption spectra in Fig.~7 with experimental data on Fig.~1 for a given delay. The optical pumping is set 30 meV above the band edge, in the continuum, thus creating an initial unbound electron-hole plasma.

\begin{figure}[!rh]
\centering
\scalebox{0.33}{\includegraphics*{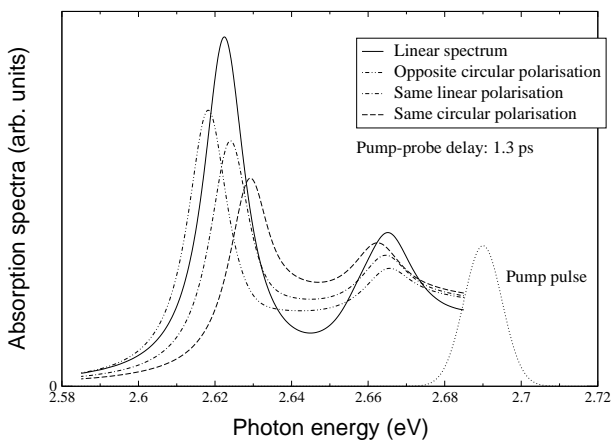}}
\caption{Calculated absorption spectra. Comparison of OCP, SLP and SCP absorption.}
\end{figure}

The experimental and numerical spectra look similar, but one can observe a significant redshift of the OCP exciton peak that is not observed on experimental data. This artificial redshift is mainly due to the screened Hartree-Fock approximation \cite{JAH96} and also to the screening model leading to an overestimation of the bandgap renormalization \cite{ELL89}. The heavy-hole exciton peak is more bleached in SCP than it is in OCP, and it is more blue shifted. This is due to the phase space filling effect that is more important in the SCP configuration. For the light-hole exciton, as in the experiment, we observe the opposite phenomenon: the OCP light-hole exciton peak is more bleached and blue shifted than the SCP light hole exciton peak. This is because the spin-up electrons excited with the $\sigma^-$- polarized pump, from the heavy hole transition, occupy states that would be created from a light-hole transition with the $\sigma^+$-polarized pump. So, in the OCP situation, the reduction of the oscillator strength owing to filling of the phase space when one probes the light-hole transition is more important than it is for SCP. The SLP configuration can be seen as an intermediate case between OCP and SCP.

\subsection{Heavy hole exciton peak dynamics}

As we mentioned above, we are concerned with the time evolution of the absorption spectra. So we computed both the bleaching and the energy shift of the exciton resonances as functions of the time delay between the pump and the probe. The numerical results in Fig.~8 show a decay of the heavy-hole exciton peak bleachings and energy shifts for OCP, SLP and SCP. The three curves converge after a few tens of picoseconds, because of spin-flip and recombination.\\

\begin{figure}[!rh]
\centering
\scalebox{0.33}{\includegraphics*{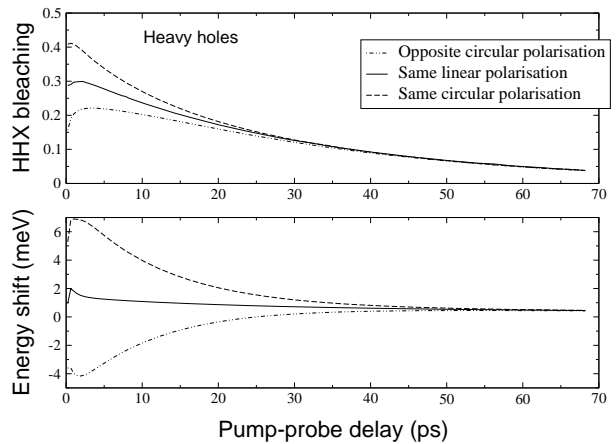}}
\caption{Calculated absorption spectra dynamics. Comparison of OCP, SLP and SCP heavy-hole exciton peaks' bleaching and shift.}
\end{figure}

{\bf OCP}: The initial state of the electron–hole plasma given in Fig.~3 shows that the only contribution to heavy-hole exciton bleaching is due to plasma screening, which lowers the Coulomb enhancement. The rapid equilibration of the carrier distributions, together with plasma cooling, contributes to the increase of bleaching at early times, i.e., less than 5 ps. Although in this case the carrier spin-flip occurs on the same time scale as the recombination, the latter process is dominant, and, with decreasing plasma density plasma screening and phase-space filling become less important. Thus the amount of bleaching decreases. In terms of energy shift, the initial heavy-hole exciton peak is redshifted. The bandgap renormalization is not strong enough to compensate for the exciton binding energy, which remains large because of the absence of the Pauli blocking effect. The redshift becomes even more important as the plasma screening is increased, because of the fast light-hole scattering. Then, because of the carrier spin flip and the recombination process, the heavy-hole exciton peak shifts toward the blue region and saturates on a longer time scale (from 30 ps). This behavior is qualitatively different from what we observe in the experiments because of the initial large redshift: If the initial heavy-hole exciton peak were not redshifted it would enter the blue region because of the increasing phase-space filling effect owing to the spin flip and to plasma cooling; then, because of recombination, we would observe a shift toward the red region, which would explain the presence of a maximum.

{\bf SCP}: As well as plasma screening, phase-space filling contributes to bleaching of the heavy-hole exciton peak. Because of thermalization and plasma cooling, we observe a slight increase for short ($\le$ 2ps) pump--probe delays, but the carrier spin flip decreases the amount of bleaching together with the recombination process. In SCP both processes contribute to decreasing the Pauli blocking filling effect. The initial SCP heavy-hole exciton peak is blueshifted, even though the bandgap renormalization is more important than it is for OCP (the exchange term is nonzero). This is so because the phase-space filling factor decreases the exciton binding energy significantly in the SCP case. Because of the fast relaxation of the distributions and the plasma cooling, the heavy-hole exciton blueshift is even more important at early times ($\le$ 2ps); then the spin flip and recombination make the heavy-hole exciton peak shift toward the red region. In the SCP situation, carrier spin–flip and recombination have the same effects on the exciton peak dynamics, whereas in the OCP case they tend to cancel each other out.

{\bf SLP}: As for SCP, plasma screening and phase-space filling contribute to heavy-hole exciton peak bleaching. There is an initial small increase that is due to thermalization and plasma cooling, but after a few picoseconds the recombination process starts effectively to influence the bleaching behavior. In SLP the spin-flip process plays no role in the dynamics of the bleaching. The behavior of the energy shift for SLP is similar to but less dramatic than the OCP behavior, as it is not influenced by the spin-flip process.

\subsection{Light hole exciton peak dynamics}

In light hole transitions the contribution of the light hole population becomes negligible quickly as the light hole density decays on a vey short time scale, $\tau_{lh} = 0.5$ ps. The behavior of the light-hole exciton peak bleaching and shift is dominated by the electrons. As we mentioned above, the bleaching here is more important for OCP than for SCP. Because of the rapid light-hole density decay, the bleaching decreases at very early times (less than 1 ps), but it increases shortly after that because of electron thermalization and plasma cooling, as shown in Fig.~9. Then, after a few picoseconds, the recombination process dominates the behavior of the bleaching, which decreases monotonically with increasing pump–probe delay. The three curves converge in $\sim$30 ps. As far as the energy shift is concerned, the blueshift quickly becomes small because of the fast light-hole density decay.

\begin{figure}[!rh]
\centering
\scalebox{0.33}{\includegraphics*{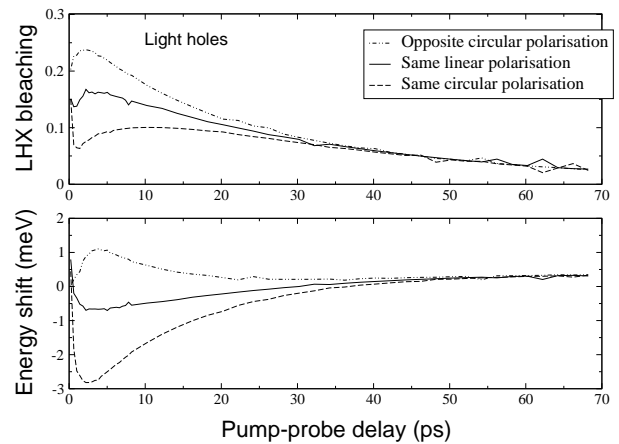}}
\caption{Calculated absorption spectra dynamics. Comparison of OCP, SLP and SCP light-hole exciton peaks' bleaching and shift.}
\end{figure}

Calculations made with longer characteristic times for the plasma thermalization and cooling ($\ge$ 5 ps) led to results that are qualitatively quite different from exprimental data: In Fig.~2 the time evolution of the heavy-hole exciton bleaching is characterized by a simple decay in OCP and SCP, whereas our calculations show the presence of a maximum for short pump–probe delay that is due to increased Pauli blocking because of plasma cooling. If the cooling characteristic time is increased, the numerical results show a plateaulike behavior for pump–probe delays of as much as several picoseconds before the recombination becomes dominant. The early ($\le$ 5 ps) behavior shown in Fig.~8 for OCP is enhanced, and the SCP bleaching dynamics is qualitatively the same as that of OCP.

\section{Discussion and conclusion}

We have presented a model that describes the time evolution of an initial hot electron-hole plasma and its influence on absorption spectra. We considered spin-selective excitation and included various dynamical process in our model. Except for the energy shift of the OCP heavy-hole exciton, the numerical results are in good qualitative agreement with the experiment. According to our calculations, the electron spin populations equilibrate on a much longer time scale (30 ps) than the thermalization of the electron-hole plasma (1 ps).

This spin-flip time is consistent with the experimental values for the II$-$VI quantum wells discussed in Ref.~\cite{HAG99}. One would need more-extensive experimental data to obtain more conclusive results for the heavy-and light-hole spin-flip times. We made additional numerical calculations to explore various spin-flip time regimes. The results show that, when the electron spin-flip time is short (1 ps), the OCP and SCP curves converge faster than when it is long, even with a short heavy-hole spin-flip time. This means that the electrons, which are lighter than the heavy holes, always dominate phase-space filling, even if the initial electron gas temperature is much higher than the heavy hole gas's temperature (Fig. 6). This implies that our results are rather insensitive to the specific value of the heavy-hole spin flip time. The light holes do not play a siginifcant role in phase-space filling, as they decay on a very short time scale (0.5 ps). Plasma cooling enhances plasma screening, and together with the relaxation of the distributions it also enhances the Pauli blocking effect. We can observe the influence of these fast processes on the early behavior of the exciton peaks' bleaching and energy shift. The spin-flip process leads to opposite qualitative behavior of the bleaching dynamics, depending on the initial polarization configurations. As shown above, exciton peak bleaching and shift can either increase or decrease because of the spin-flip process, depending on the type of exciton (heavy or light) and on the polarization (OCP or SCP). The radiative recombination occurs on a time scale that is too large ($\sim$1.6 ns in ZnSe) to have any effect on the fast bleaching and energy shift dynamics (below 100 ps). However, the nonradiative recombination is fast enough that an overall decay can be observed on a time scale shorter than 100 ps.

As far as the energy shift is concerned, the model that we used for this study seems not to be good enough to describe qualitatively the OCP and SCP energy shifts. We make two comments on this problem: First, we have restricted ourselves to the screened Hartree-Fock level, neglecting additional correlation terms to account for broadening, which is described here by phenomenological damping. Inclusion of such terms has been shown to correct spurious shifts introduced by the screened Hartree-Fock approximation \cite{JAH96} in SLP. However, the static plasmon-pole approximation that we and many others have used is constructed in just such a way as to produce a constant exciton energy at the screened Hartree-Fock level in SLP. We do this by adding a term to the effective plasmon frequency to account for the effects of the pair continuum. As can be seen from Fig. 7, there is effectively no shift of SLP in our calculations; we also checked our numerics against published GaAs results. Making the natural extension to OCP and SCP, however, leads to extraneous shifts, even when we include this pair continuum term. Our conclusion is therefore that the static plasmon-pole approximation is inadequate for the treatment of spin-polarized populations. Of course one could alter the strength of the pair-continuum term in an \emph{ad hoc} fashion to correct for this, but that correction would accomplish little. Within our model it is possible to control the values of the phenomenological parameters to obtain further insight into the complex interplay among various dynamical processes as more experimental data become available.

\end{document}